\documentclass[runningheads,a4paper]{llncs}

\usepackage{amssymb,amsmath,makeidx,enumerate,stmaryrd,textcomp}
\usepackage[all]{xy}
\usepackage{tabls}
\usepackage{pslatex}

\usepackage{color,graphicx,xcolor}

\newif\ifcscomp
  \cscomptrue
  \cscompfalse

\newif\ifcost
  \costfalse
  \costtrue

\newif\ifnat
  \natfalse
  \nattrue

\newif\ifemi
  \emitrue
  \emifalse

\newif\ifthm
  \thmfalse
  \thmtrue

\newif\ifllncs
  \llncsfalse
  \llncstrue

\newif\ifarticle
  \articletrue
  \articlefalse

\newif\iftesi
  \tesitrue
  \tesifalse

\newif\ifmod
  \modtrue
  \modfalse

\newif\ifqapl
  \qapltrue
  \qaplfalse

\newif\ifqapl
  \qaplfalse
  \qapltrue

\newif\ifpar
  \partrue
  \parfalse

\newcommand{\hdof}[1]{\mathcal H_{\nretohds{#1}}}
\newcommand{\nrebinder}[1]{\tuple{#1}}
\newcommand{\q}{\mathsf{q}}
\newcommand{\x}{\mathsf{x}}
\newcommand{\y}{\mathsf{y}}

\newcommand{\longversion}[1]{#1}
\newcommand{\oscope}{\mbox{\tiny$\langle\!\!\langle$}}
\newcommand{\cscope}{\mbox{\tiny$\rangle\!\!\rangle$}}
\newcommand{\nretohds}[1]{\llparenthesis #1 \rrparenthesis}
\newcommand{\mybinder}[2]{\oscope {#1} . {#2} \cscope}
\newcommand{\addname}[2]{#1 \dag  #2}

\newcommand{\alp}{\mathcal S} 

\newcommand{\stk}{\Sigma}
\newcommand{\pushtr}{\curvearrowright}
\newcommand{\poptr}{\curvearrowleft}
\newcommand{\popstk}[1]{\stackrel{\poptr}{#1}}
\newcommand{\popstktwo}[1]{\stackrel{2\poptr}{#1}}
\newcommand{\topstk}[1]{{#1}^\top}
\newcommand{\stkupd}[2]{{#1} \bullet {#2}}
\def\finex{{\unskip\nobreak\hfil
\penalty50\hskip1em\null\nobreak\hfil$\diamond$
\parfillskip=0pt\finalhyphendemerits=0\endgraf}}

\usepackage{fge}

\newcommand{\estk}{\fgestruckzero}

\newcommand{\step}[3]{#2 \stackrel{#1}{\to} #3}

\newcommand{\trans}{\mathit{tr}}

\newcommand{\rec}{\mathit{rec}}
\newcommand{\emptystr}{\epsilon}

\newcommand{\lang}[1]{\mathcal{L}_{#1}}

\newcommand{\fresh}{\star}
\newcommand{\pop}[1]{\mathtt{pop}}
\newcommand{\push}[1]{\mathtt{push}}

\newcommand{\weight}[1]{|#1|}

 \newcommand{\Real}[1]{\mathrm{Real}}

  \newcommand{\compile}[2]{\ifthenelse{\equal{#1}{yes}}{#2}{}}

  \newcommand{\cf}[2]{
    \fontsize{#1}{#1}{\selectfont{#2}}
  }
  \ifemi
  \usepackage{ulem}
    \newcommand{\emi}[1]{{\marginpar{\cf{6}{{#1}}}}}
    \newcommand{\emic}[2]{\ \\[.1em]
      \definecolor{shadecolor}{rgb}{1,0.99,0.9}
      \fcolorbox{red}{shadecolor}{\parbox{\linewidth}{ 
            \color{gray}
            {\color{blue} #2}{\sf #1}
            }}
      \ \\[.1em]}
    
    \else
    \newcommand{\emi}[1]{}
    \newcommand{\emic}[2]{}
    
    \fi

  \ifemi

  \else

  \fi

  \newcommand{\st}{\ \ \big| \ \ }

     \ifmod
       
     \else
       
     \fi

  \ifcscomp
    \newcommand{\comp}[2]{#1 ; #2}
  \else
    \newcommand{\comp}[2]{#2 \circ #1}
  \fi
  \newcommand{\id}[1]{\mathrm{id}_{#1}}

  \newcommand{\proofend}{\mbox{$\Box$}}

  \newcommand{\mmdef}{\mbox{$\;\;\stackrel{\mathrm{def}}{=}\;\;$}}

  \ifnat
  \fi

  \newcommand{\supp}{\mathit{supp}}

  \newcommand{\upd}[2]{[{#1} \mapsto {#2}]}

  \newcommand{\hda}{HDA}
  \newcommand{\hds}{HDS}

  \newcommand{\hdns}{HDS}

  \ifpar
    
  \else
    
  \fi

  \newcommand{\enc}[2]{\{#1\}_{#2}}

  \newcommand{\strans}[1]{\xymatrix{\ar@{~>}[r]^{#1} &}}
  \newcommand{\dstrans}[1]{\xymatrix{\ar@{~>}[d]^{#1} \\ \ }}

  \newcommand{\comment}[1]{}
  \iftesi

  \else

  \fi

\newcommand{\conf}[1]{\langle {#1} \rangle}

\iftesi
  
\else
  
\fi

\newcommand{\names}{\mbox{$\mathcal{N}$}}
\newcommand{\namestar}{\mbox{$\mathcal{N^\ast}$}}

\ifarticle
  \newtheorem{theorem}{Theorem}[section]
  \newtheorem{definition}{Theorem}[section] 
  \newtheorem{proposition}{Theorem}[section] 
  \newtheorem{lemma}{Theorem}[section]
  \newtheorem{corollary}{Theorem}[lemma] 
  \newtheorem{remark}{Theorem}[section]
  \newtheorem{observation}{Theorem}[section]
  \newtheorem{notation}{Theorem}[section]
  \newtheorem{example}{Theorem}[section]
\else
  \ifthm
  \else
    \newtheorem{theorem}{Theorem}[section]
    \newtheorem{definition}{Definition}[section] 
    \newtheorem{proposition}{Proposition}[section] 
    \newtheorem{lemma}{Lemma}[section]
     
    \newtheorem{remark}{Remark}[section]

    \newtheorem{example}{Example}[section]
  \fi
\fi

\ifpar
\else

\fi

\newcommand{\fpart}[1]{\ensuremath{\mathcal{P}_{\omega}\ {#1}}}

\iftesi
  
\else
  
\fi

\newcommand{\Q}{{\cal Q}}

\newcommand{\dom}[1]{\mathit{dom}(#1)}
\newcommand{\cod}[1]{\mathit{cod}(#1)}

\newcommand{\tuple}[1]{\langle#1\rangle}
\newcommand{\hdtr}[3]{\tuple{#2,#1,#3}}

\usepackage{url}
\urldef{\mailsa}\path|{kurz, tomoyuki.suzuki, emilio}@mcs.le.ac.uk|

\begin{document}

\mainmatter  

\title{Towards Nominal Formal Languages \\ (long version)}

\titlerunning{Towards Nominal Formal Languages}

%
%
\author{Alexander Kurz \and Tomoyuki Suzuki\thanks{The author's PhD research is supported by Yoshida Scholarship Foundation.} \and Emilio Tuosto}
\authorrunning{A.~Kurz, T.~Suzuki and E.~Tuosto}

\institute{Department of Computer Science, University of Leicester, UK
}

%
%

\toctitle{Lecture Notes in Computer Science}
\tocauthor{Authors' Instructions}
\maketitle

\begin{abstract}
  We introduce formal languages over infinite alphabets where words
  may contain binders. We define the notions of nominal language,
  nominal monoid, and nominal regular expressions.  Moreover, we
  extend history-dependent automata (HD-automata) by adding stack, and
  study the recognisability of nominal languages.
  \comment{
  Finally, we show how our framework can be used to model
  correctness of a cryptographic protocol models in a $\pi$-like
  calculus.
}

\end{abstract}

\section{Introduction}

Automata over infinite alphabets have been receiving an increasing
amount of attention, see eg
\cite{kaminskifrancez94,segoufin:csl06,bojanczyk:stacs11,tze11}. In
these approaches, the countably infinite alphabet $\names$ can be
considered as a set of `names', which can be tested only for
equality. Typically, languages of interest such as
\begin{equation}\label{exle:lang-kf1}
\mathcal{L}_1 = \{n_1\ldots n_k \in\namestar\mid \exists i \not= j\,.\,  n_i=n_j\}
\end{equation}
from \cite{kaminskifrancez94} are invariant under name-permutations:
If eg $nmn$ is in the language, then so is $n'mn'=(n\ n')\cdot nmn$,
where $(n\ n')\cdot nmn$ stands for the application of the transposition
$(n\ n')$ to the word $nmn$. This suggests to think of the names as
being bound and languages to be closed under $\alpha$-equivalence. On
the other hand, we may fix a name $n_1$ and consider the language
\begin{equation}\label{exle:lang-kf2}
  \mathcal{L}_{2,n_1} = \{n_1 n_2\ldots n_k \in\namestar \mid \forall
  i\not=j\,.\, n_i\not= n_j\}
\end{equation}
from \cite{tze11}; we can think of $n_1$ as a free name and of the
$n_2, \ldots n_k$ as bound. This suggests to study not only words over
names, but also words which contain binders and allow us to make
explicit the distinction between bound and free names. Automata on
words with binders already appear in \cite{stirling:fossacs09} in the
study of the $\lambda$-calculus. \emph{In this paper we begin the
  systematic study of words with binders from the point of view of the
  classical theory of formal languages and automata.}

\medskip\noindent In particular, our contributions are:
\begin{itemize}
\item \emph{nominal languages} of words with binders
  (\S~\ref{sec:nomlang}) as a natural generalisation of formal
  languages over infinite alphabets;
\item \emph{nominal monoids} (\S~\ref{sec:nom-monoid}) as the
  corresponding algebraic structures;
\item \emph{nominal regular expressions} (\S~\ref{nomre:sec}) as a
  generalisation of regular expressions;
\item \emph{HD-automata with stack} (\hdns) (\S~\ref{sec:hdns}) and
  Theorem~\ref{thm:hdnsnre} showing that nominal regular expressions
  can be faithfully encoded into \hdns.
\end{itemize}

\noindent One of the motivations to study words with binders comes
from verification. For instance, consider the Needham-Schroeder
protocol
\[\begin{array}{l}
 A \to B: \enc{n, A} B \\
 B \to A: \enc{n, m} A \\
 A \to B: \enc m B
\end{array}\]
The (correct) runs of the protocol can be characterised by a nominal
regular expression
\begin{equation}\label{equ:ns}
 \langle n . \ \mathtt{ENCR}\ n\ A\ \mathtt{FOR}\ B \ \
 \langle m . \ \mathtt{ENCR}\ n\ m\ \mathtt{FOR}\ A \ \
 (\mathtt{ENCR}\ m\ \mathtt{FOR}\ B)\rangle\rangle^\ast
\end{equation}
where the alphabet is now $\names\cup\mathcal{S}$ with $n,m\in\names$
and $\{\mathtt{ENCR},\mathtt{FOR},A,B\}=\mathcal{S}$ a finite set of
`letters'; finally, $\langle n.e \rangle$ binds all the free
occurrences of $n$ in $e$ and \emph{generates a fresh name} $n$.
From~(\ref{equ:ns}) one could obtain an \hdns\ for monitoring the
execution of a protocol, i.e.\ the \hdns\ would be able to detect if
something goes wrong during the execution (e.g., an intruder is
performing an attack). From an automata theoretic point of view, the
interesting new feature appears more clearly if we abstract
(\ref{equ:ns}) to
\begin{equation}\label{equ:ns2}
 \langle n . n \langle m . nm \rangle\rangle^\ast
\end{equation}
and note that binding (fresh name generation) $\langle\_.\_\rangle$
appears under the Kleene star, which is the reason why automata
accepting such languages need to have a stack.

\section{Nominal Languages}\label{sec:nomlang}
\newcommand{\letters}{\mathcal{S}}
\newcommand{\mwords}{\mathbf{M}}
\newcommand{\gwords}{\mathbf{G}}
\newcommand{\lwords}{\mathbf{L}}
\newcommand{\swords}{\mathbf{S}}
\newcommand{\sltr}{\mathrm{sl}}
\newcommand{\lgtr}{\mathrm{lg}}
\newcommand{\gmtr}{\mathrm{gm}}
\newcommand{\mpwtr}{f_\mwords}

We introduce languages with name binders. This section appeals to our
intuitive understanding of binding and $\alpha$-equivalence as known
from eg $\lambda$-calculus or first-order logic, but see the next
section for a formal treatment.
To start with, the \emph{alphabet} is divided disjointly into a
countably infinite set $\names$ (of \emph{names}) and a finite set
$\mathcal{S}$ (of \emph{letters}).

\begin{definition}[m-word]
  An \emph{m-word} is a term built from constants
  $\names\cup\mathcal{S}\cup\{\emptystr\}$, and two binary operations
  $\circ,\oscope\_ .\_ \cscope$, according to
$$w \mmdef \emptystr \mid n \mid s \mid w{\circ}w \mid \mybinder{n}{w},$$
where $n$ ranges over $\names$ and $s$ over $\letters$. We denote by
$\mwords$ the set of all m-words.
\end{definition}
As in the classical case we assume that $\emptystr$ (the empty word)
is the neutral element wrt $\circ$ and that $\circ$ is associative. We
often write $wv$ for the concatenation $w\circ v$. Furthermore, we let
$\mybinder{n}{w}$ bind the free occurrences of $n$ in $w$ and take
m-words up to $\alpha$-equivalence.

The notion of m-word is the \emph{m}ost general notion of word with
binders: We only require from words to form a monoid and behave well
wrt $\alpha$-equivalence. Due to the scope introduced by binding,
words now have a tree structure. This motivates the following, more
special, but perhaps more naturally \emph{g}eneralised, notion of
words.
\begin{definition}[g-word]
  A \emph{g-word} is a term built from $\emptystr$, unary operations
  $n\_$, $s\_$ for each $n\in\names,s\in\letters$, and a binary operation
  $\oscope\_ .\_ \cscope$, according to
$$w \mmdef \emptystr \mid nw \mid sw \mid \mybinder{n}{w}.$$
We denote by $\gwords$ the set of all g-words.
\end{definition}
Regarding binding and $\alpha$-equivalence, we follow the same
conventions as for m-words. To consider $\gwords$ as a monoid, we
define $\circ\colon \gwords\times \gwords\to\gwords$ as
follows:
\begin{equation}\label{equ:g-concat}
\begin{array}{lcl}
  \emptystr{\circ}w \mmdef w &\quad\quad\quad\quad&
  nw{\circ}v \mmdef n(w{\circ}v)\\
  sw{\circ}v \mmdef s(w{\circ}v) &&
  \mybinder{n}{w}{\circ}v \mmdef \mybinder{n'}{(w'{\circ}v)}
\end{array}
\end{equation}
where $n'$ is fresh for $v$ and $\mybinder{n'}{w'}$ is an
$\alpha$-renaming of $\mybinder{n}{w}$.
Intuitively speaking, we extrude the scope of the binding to the end
of the word.

Next we allow binders to appear only at the beginning of a word.
\begin{definition}[l-word]
  An \emph{l-word} is a pair $(p,w)$ where $p\in\namestar$ and
  $w\in(\names\cup\mathcal{S})^\ast$. We denote by $\lwords$ the set
  of all l-words.
\end{definition}
We interpret $p$ as a prefix of name binders and $w$ as the part of
the word that has no binders.
$\circ\colon \lwords\times\lwords\to\lwords$ is given
on the left below 
\begin{equation}\label{equ:l-concat-binding}
(p,w){\circ}(q,v) \mmdef (pq,wv)\quad\quad\quad\quad [n](p,w) \mmdef (np,w)
\end{equation}
where we assume that $p$ and $q$, $p$ and $v$, and $q$ and $w$ have no
names in common.
Whereas previously name-binding was built into the syntax via $\oscope
\_ .\_ \cscope$, we now define explicitly, anticipating notation from
\S~\ref{sec:nom-monoid}, a binding operation $[\_]\_\colon
\names\times\lwords\to\lwords$ via the clause on the right of
\eqref{equ:l-concat-binding}.

\begin{definition}[s-word]
  An \emph{s-word} is a pair $(S,w)$ where
  $w\in(\names{\cup}\mathcal{S})^\ast$ and $S$ is a subset of the
  names appearing in $w$. We denote by $\swords$ the set of all
  s-words.
\end{definition}
On $\swords$, we define the two operations $\circ$ and $[\_]\_$ as
follows, assuming that $S$ and $T$, $S$ and $v$, $T$ and $w$ have no
names in common.
\begin{equation}\label{equ:s-concat-binding}
\begin{array}{ccc}
(S,w){\circ}(T,v) \mmdef (S\cup T, wv) &\quad\quad\quad\quad &
  [n](S,w) \mmdef \begin{cases}(S\cup\{n\},w) & \text{if $n$ 
      in $w$}\\ (S,w) & \text{otherwise}\end{cases}
\end{array}
\end{equation}

\begin{remark}\label{rmk:slgm}
We have embeddings 
$\swords\stackrel{\sltr}{\to}
\lwords\stackrel{\lgtr}{\to}
\gwords\stackrel{\gmtr}{\to}
\mwords$.
For $\sltr$ we assume that names are ordered; the other main clauses
are $\lgtr(np,w)=\oscope n.\lgtr(p,w)\cscope$ and
$\gmtr(nw)=n{\circ}\gmtr(w)$.
\end{remark}

\section{Nominal monoids}\label{sec:nom-monoid}

The somewhat informal treatment of \S~\ref{sec:nomlang} should be
sufficient to understand how automata process words with binders in
\S~\ref{sec:hdns} and \S~\ref{sec:hsnre}. On the other hand, from a
conceptual point of view, it is important to have a unifying
account. The presence of names and binders suggests to employ nominal
sets \cite{gp02}. This not only provides us with a mathematical
theory, but also a clear conceptual guidance: Follow the classical
universal algebraic account of languages and automata, but replace
sets by nominal sets. Here, we apply this to languages and monoids.

Nominal sets and their logics come in different versions. We follow
\cite{gabb-math:nom-alg}, for which we need to refer to for
details. More details can also be found in \cite{kst10}. Let us just
recall

\begin{definition}[Nominal set]
  Denote by $\mathit{Perm}(\names)$ the group of permutations of
  $\names$ generated from the set of transpositions $\{(n\ m)\mid
  n,m\in\names\}$. A set $A$ equipped with a
  $\mathit{Perm}(\names)$-action $\mathit{Perm}(\names)\times
  A\stackrel{\cdot}{\longrightarrow} A$ is a \emph{nominal set}, if
  every element in $A$ is finitely supported. This means that for each
  $a\in A$ there is finite set $S\subseteq\names$ (called a support of
  $a$) such that $\ \pi|_S=\id \ \Rightarrow \ \pi\cdot a = a\ $ for all
  $\pi\in\mathit{Perm}(\names)$ (where $\pi|_S$ denotes the
  restriction of $\pi$ to $S$). Maps between nominal sets are required
  to be equivariant, that is, they respect the permutation action.
\end{definition}
It follows 
that each element $a\in A$ has a minimal support $\supp(a)$ and one
writes $\mathbf{n\#a}$ ($n$ is \emph{fresh} for $a$) for
$n\notin\supp(a)$. This allows us to define abstraction \cite[Lemma
5.1]{gp02} as $[n]a\mmdef\{(n,a)\}\cup\{(m,(nm)\cdot a) \mid m\#a\}$
and 
$[\names]A\mmdef\{[n]a\mid n\in\names, a\in A\}$.

A \textbf{nominal algebra} $\mathfrak{A}$, see \cite[Def
4.13]{gabb-math:nom-alg}, consists of a nominal set $A$, constants
$n\in\names$, and a map $[\names]A\to A$. As in universal algebra,
further operations and equations may be added:
\begin{definition}
  A \emph{nominal monoid} is a nominal algebra $\mathfrak{A}$ with
  additional constants $s\in\letters$ and (equivariant) operations
  $\emptystr,\circ$ so that $(A,\emptystr,\circ)$ is a monoid.
\end{definition}
We say that $w\in A$ is \textbf{closed}, or that $w$ contains no free
names, if $\supp(w)$ is empty. 
\begin{definition}
  Write $\mathcal{C}_m$ for the class of all nominal monoids.  We
  consider the following axioms where $m,n\in\names$, $s\in\letters$,
  and $X,Y$ are variables ranging over carriers of algebras.
  \begin{equation*}
    \begin{array}{llcll}
\textbf{Ax1 \quad} & n\# Y \vdash [n]X{\circ}Y = [n](X{\circ}Y) 
&\quad\quad&
\textbf{Ax2} & \vdash s{\circ}[m]Y = [m](s{\circ}Y) \\
\textbf{Ax3} & n\# m \vdash n{\circ}[m]Y = [m](n{\circ}Y)
&&
\textbf{Ax4} & \vdash [n][m]X = [m][n]X \\
\textbf{Ax5} & n\# X \vdash [n]X = X
&&
\textbf{Ax6} & n\# X \vdash X{\circ}[n]Y = [n](X{\circ}Y)
    \end{array}
  \end{equation*}     
  $\mathcal{C}_{g}$,
  $\mathcal{C}_l$, $\mathcal{C}_s$ are axiomatised by Ax1, Ax1-3,
  Ax1-5, respectively.
\end{definition}

\begin{remark}\label{rmk:axioms}
  One possible reading of the operations and the axioms is as
  follows. In $\mwords$, we have sequential composition $\circ$,
  allocation $\oscope n$ of a resource named $n$, and deallocation
  $\cscope$. In $\gwords$, we don't care about deallocation (garbage
  collection). In $\lwords$, the timing of the allocation does not
  matter and all resources may be allocated at the start. In
  $\swords$, the order of allocation does not matter and the
  allocation of an unused resource is redundant.

  But other interpretations are possible. With $[n]$ as the $\nu n$ of
  the $\pi$-calculus and $\circ$ as $\mid$, Ax6 becomes the familiar
  law of scope extrusion. Interpreting $[n]$ as $\forall$, Ax4-5 are
  familiar laws of the universal quantifier. In \cite{pitts:popl10}, a
  binder satisfying Ax4-5 is called a name-restriction operator.
\end{remark}
We can now summarise the previous section conveniently in
Table~\ref{tb:axioms} and
\begin{theorem}\label{thm:monoids}
  $\mwords,\gwords,\lwords,\swords$ are the initial monoids in,
  respectively, $\mathcal{C}_m$, $\mathcal{C}_{g}$, $\mathcal{C}_l$
  and $\mathcal{C}_s$.
\end{theorem}
\begin{proof}
  The detailed proof can be found in~\cite{kst10}.
  \qed
\end{proof}
\begin{table}[t]
\begin{center}
 \caption{Summary of nominal monoids and the axioms}
 \label{tb:axioms}
\begin{tabular}{|c|p{4cm}|c|p{3.6cm}|}
  \hline
  Classes & Axioms & Initial monoid & Typical example \\
  \hline 
  $\mathcal{C}_m$ & & $\mathbf{M}$ & $[n_1](s_1n_1n_4)[n_0](n_0[n_3]s_2)$  \\
  \hline
  $\mathcal{C}_{g}$ & Ax1 &  
  $\mathbf{G}$ & $[n_1](s_1n_1n_4[n_0](n_0[n_3]s_2))$ \\
  \hline
  $\mathcal{C}_{l}$& Ax1-3 
                   & $\mathbf{L}$ 
                   & $[n_1][n_0][n_3]s_1n_1n_4n_0s_2$  \\
  \hline
  $\mathcal{C}_{s}$& Ax1-5
                   & $\mathbf{S}$ 
                   & $[n_0][n_1]s_1n_1n_4n_0s_2$\\
  \hline
\end{tabular}
\end{center}
\end{table}
\begin{remark}\label{rmk:plain-words}
  We have a mapping $\mpwtr:2^{\mwords}\to
  2^{(\mathcal{N}{\cup}\letters)^\ast}$ to
  \textbf{plain words} (ie words without binders) determined by
$\mpwtr(\{\oscope n.w\cscope\}) = \mpwtr(\{w\}) \cup \{(n\ m) \cdot v \mid v\in \mpwtr(\{w\}), m\# v\}$.
With the embedding $\gmtr\circ\lgtr\circ\sltr$ from
Remark~\ref{rmk:slgm} this induces a map $f_{\swords}$ from languages
of s-words to subsets of $(\mathcal{N}{\cup}\letters)^\ast$, eg
$f_{\swords}(\{(\{n\},n)\})=f_\mwords(\{\oscope n.n\cscope\})=\names$.
\end{remark}

\section{Nominal Regular Expressions}\label{nomre:sec}

In analogy to the classical definition, we introduce \emph{nominal
  regular expressions}:
\begin{equation}\label{eq:nre}
 e ::= 1 \mid 0 \mid n \mid s \mid e + e \mid e{\circ}e \mid \langle n.e\rangle \mid e^\ast
\end{equation}
where $n \in \names$ and $s \in \mathcal{S}$.
The semantic interpretation $L$ is defined as follows.
\begin{enumerate}
 \item $L(1) \mmdef \{\emptystr\}$, \qquad $L(0) \mmdef \emptyset$, \qquad $L(n) \mmdef \{n\}$, \qquad $L(s) \mmdef \{s\}$,
 \item $L(e_1 + e_2) \mmdef L(e_1) \cup L(e_2)$,
 \item $L(e_1{\circ}e_2) \mmdef L(e_1) \circ L(e_2) \mmdef \{w_1\circ
   w_2\mid w_1\in L(e_1), w_2\in L(e_2)\}$,
 \item $L(\langle n.e \rangle) \mmdef [n]L(e)\mmdef \{[n]w\mid w\in L(e)\}$.
 \item $L(e^\ast) \mmdef \displaystyle\bigcup_{i \in \mathbb{N}}L(e)^i$, where $L(e)^i \mmdef \underbrace{L(e) \circ \cdots \circ L(e)}_{i \text{times}}$,
\end{enumerate}
    \begin{remark}\label{rmk:nre}
      The definitions of $\circ$ and $[\_]\_$ are dependent on the
      choice of row in Table~\ref{tb:axioms}, compare
      \eqref{equ:g-concat}, \eqref{equ:l-concat-binding},
      \eqref{equ:s-concat-binding}.
      For example, on $\mwords$ we have
      $[n]L(e) = \{\mybinder{n}{w} \mid w \in L(e)\}$
and on $\lwords$ we have
$[n]L(e) = \{(np,w) \mid (p,w) \in L(e)\}$. From \S~\ref{sec:hdns}
onwards, we will interpret regular expressions in $\mwords$ only.
    \end{remark}

\begin{example}\label{exle:lang-bind}
  We have seen in \eqref{equ:ns2} how $\langle n . n \langle m . nm
  \rangle\rangle^\ast$ arises from the Needham-Schroeder protocol.
  In \S~\ref{sec:hsnre} we consider the simpler expression
  $m(\nrebinder{n.mn})^\ast$ which intuitively represent the
  computations of a security protocol where (an unbound number of) new
  `nonces' $n$ are generated within a session $m$ and should always be
  paired up with $m$.
  \finex
\end{example}

We can also interpret nominal regular expressions in plain words. For
example, let $L$ take values in $\swords$ and let
$f_{\swords}:2^{\swords}\to 2^{(\mathcal{N}{\cup}\mathcal{S})^\ast}$
denote the map of Remark~\ref{rmk:plain-words}.

\begin{example}\label{exle:tzev-lang}
  If we interpret $\langle n. n \rangle^\ast$ in $\swords$ 
  (or $\lwords$ or $\gwords$) 
  we obtain the language
\begin{equation*}\label{eqn:lang2} 
\mathcal{L}_2
\mmdef f_{\swords}(L(\langle n. n  \rangle^\ast)) 
= \{n_1 \cdots
  n_k \mid \forall i\not=j \,.\,n_i \not=
  n_j\},
\end{equation*} 
which is the complement of $\mathcal{L}_1$ from
\eqref{exle:lang-kf1}. $\mathcal{L}_2$ is not recognised by the FMAs
of \cite{kaminskifrancez94} but it is recognised by the FRAs of
\cite{tze11}. The latter notes that $\mathcal{L}_2 \ast \mathcal{L}_2
=\{wv\mid w,v\in\mathcal{L}_2\}$ shows that languages recognised by
FRAs are not closed under composition. On the other hand, the presence
of binders allows us to use $\circ$ (respecting the 'hidden' binders)
instead of $\ast$ and we obtain $\mathcal{L}_2 \circ \mathcal{L}_2 =
f_{\swords}(L(\langle n. n \rangle^\ast)) \circ f_{\swords}(L(\langle
n. n \rangle^\ast)) \mmdef f_{\swords}(L(\langle n. n
\rangle^\ast\circ\langle n. n \rangle^\ast)) =f_{\swords}(L(\langle
n. n \rangle^\ast))=\mathcal{L}_2$, where the second equality is our
definition of `nominal concatenation' on languages of
plain words. This indicates that even for languages without binders
the composition with binders is a natural concept.

Similarly, if we interpret $e=\langle l. l{\circ}\langle m. \langle
n. m{\circ}n \rangle \rangle^\ast \rangle$ in $\mwords$
we obtain another example of Tzevelekos:
\begin{equation*}\label{eqn:lang1}
f_\mwords(L(e))
=
\{m n^1_1n^2_1n^1_2n^2_2\cdots n^1_kn^2_k \mid \forall i \in \mathbb{N}, \forall j \in \{1,2\}.\ m\not=n^j_i \& n^1_i \not= n^2_i\}
\end{equation*}
\end{example}

\section{History-dependent Automata with Stack}\label{sec:hdns}
We build our nominal automata theory on \hda\ (after
\emph{history-dependent automata})~\cite{pis99}.
\hda\ are a computational model of nominal calculi defined on the
notion of \emph{named sets} and extend classical automata with finite
sets of names \emph{local} to states and transitions.
We equip \hda\ with stack; this renders them suitable for recognising
nominal languages interpreted in $\mwords$.
We argue that \hda\ are natural candidates to build a theory of
automata of nominal languages with binders.
In fact, they are equipped with mechanisms to capture name restriction
of nominal calculi~\cite{fmt05,fmtkv04,fmt03-b} and formally linked to
the nominal set theory in~\cite{gmm06,fs:cmcs04}.

Let $\fresh \not\in \names$ be a distinguished name; a \emph{stack}
$\stk$ is a sequence of finite partial maps $\sigma: \names \to \names
\cup \{\fresh\}$ (we use $\bot$ to denote the empty map).
The empty stack is denoted by $\estk$, a stack with head $\sigma$ is
written $\sigma :: \stk$, and
\[
\popstk{\stk} \mmdef \begin{cases}
    \stk', &  \stk = \sigma::\stk'
    \\
    \estk, & \stk = \estk
  \end{cases}
\qquad
\popstktwo \stk \mmdef \popstk{(\popstk \stk)}
\qquad
\topstk{\stk} \mmdef \begin{cases}
    \sigma, &  \stk = \sigma::\stk'
    \\
    \bot, & \stk = \estk
  \end{cases}
\]
respectively are the pop, pop twice, and top operations.

\begin{definition}[\cite{pis99}]\label{def:nset}
  A \emph{(basic) named set} $\tuple{Q,\weight{\_}_Q}$ is a set $Q$
  (of states) with a map $\weight{\_}_Q: Q \to \fpart \names$ sending
  $q \in Q$ to a finite set of names $\weight{q}_Q$ (called
  \emph{local names of $q$}).
\end{definition}
\longversion{
  Basically, the elements $q$ of a named set are equipped with a set
  of \emph{local} names $\weight q_Q$.
}
Hereafter we omit subscripts when clear from the context and write a
named set $\tuple{Q,\weight{\_}_Q}$ as $Q$, in which case
$\weight{\_}$ is understood as the map of local names of $Q$; also,
the \emph{update of a map $f: X \to Y$ at $x$ with $y$} is the map
\[
f\upd x y:
X \cup \{x\} \to Y \cup \{y\}
\quad \text{ such that } \quad
(f\upd x y)(a) = \left\{\begin{array}{ll}
    f(a), & \text{if } a \neq x
    \\
    y,  & \text{if } a = x
\end{array}\right.\]

Before giving the formal definition, we intuitively present \hda\ with
stack.
A transition $\hdtr{\alpha}{q'}{\sigma}$ from a state $q$ consists of
the target state $q'$, a label $\alpha$, and a map $\sigma$ keeping
track of the correspondences of names.
Labels $\alpha$ can be a local name $n \in \weight q$ of the source
state $q$, letters $s \in \alp$, or any of the distinguished symbols
\[
  \emptystr \qquad\qquad
  \pushtr     \qquad\qquad
  \poptr      \qquad\qquad
  \oscope     \qquad\qquad
  \cscope
\]
respectively representing internal transitions, push, pop, \emph{name
  allocation}, and \emph{name deallocation}.
Example~\ref{ex:hdns} gives a convenient graphical representation of
an \hdns.
\begin{example}\label{ex:hdns}
  Let $\mathsf{q_0}$, $\mathsf q$, and $\mathsf{q'}$ be states with
  $\weight{\mathsf{q_0}} = \{x\}$, $\weight{\mathsf q} = \{z\}$, and
  $\mathsf{q'} = \emptyset$.
  The \hdns
  \begin{center}
    \begin{minipage}{\linewidth}\centering
      \includegraphics[scale=.4]{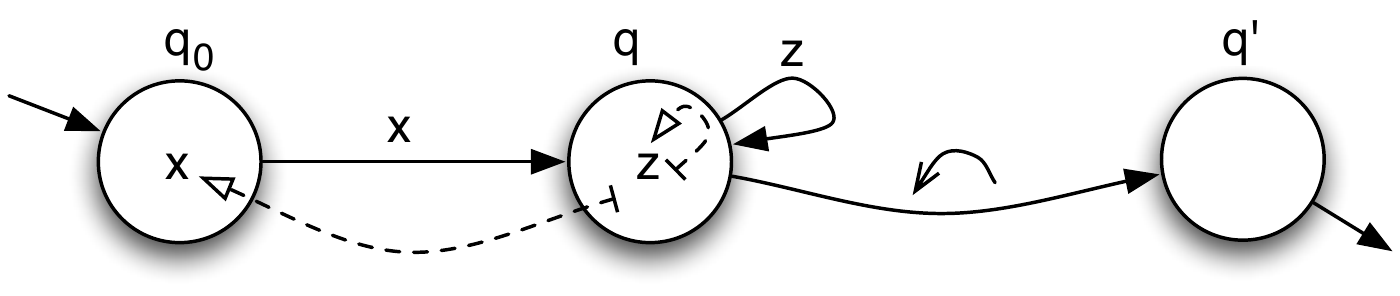}
    \end{minipage}
  \end{center}
  has initial (resp. final) state $\mathsf{q_0}$
  (resp. $\mathsf{q'}$).
  Both $\mathsf{q_0}$ and $\mathsf{q}$ have a transition exposing
  their (unique) local name ($\mathsf{x}$ and $\mathsf{z}$
  respectively).
  Maps among local names are represented by dashed arrows.
  Also, $\mathsf{q}$ has $\poptr$ transition to $\mathsf{q'}$
  with the empty map of local names.
  \finex
\end{example}

\begin{definition}\label{def:hd-automaton}
  A (non-deterministic) \emph{history-dependent automaton with stack
    on $\names \cup \alp$} (\hdns) is a tuple
  $\tuple{Q,q_0,\eta,F,\trans}$ where
  \begin{itemize}
  \item $Q$ is a named set of \emph{states} (the states of the
    automaton);
  \item $q_0 \in Q$ is the \emph{initial state};
  \item $\eta$ is a partial function from $\weight{q_0}_Q$ to
    $\names$;
  \item $F \subseteq Q$ is the named set of \emph{final states}
    with $\weight{\_}_F$ being the restriction of $\weight{\_}_Q$ to $F$;
  \item $\trans$ is the \emph{transition function} returning for each
    $q \in Q$ a finite set $\trans(q)$ of \emph{transitions}, namely
    tuples $\hdtr \alpha {q'} \sigma$ such that
    \begin{itemize}
    \item if $\alpha \in \names$ then $\alpha \in \weight{q}_Q$
    \item if $\alpha = \oscope$ then $\sigma:\weight{q'} \to
      \weight{q} \cup \{\fresh\}$
    \item if $\alpha = \pushtr$ then $\sigma:\weight{q'} \to \names$
    \item otherwise $\sigma:\weight{q'} \to \weight{q}$
    \end{itemize}
    and, in either case, $\sigma$ is a partial injective map (see
    Remark~\ref{rmk:eta} on page~\pageref{rmk:eta}).
  \end{itemize}
\end{definition}
Transitions in Def~\ref{def:hd-automaton} allow \hdns\ to accept
names or letters or to manipulate the stack.
Besides the usual \emph{push} ($\pushtr$) and \emph{pop} ($\poptr$)
operations, \hdns\ feature allocation ($\oscope$) and deallocation
($\cscope$) of names.

\begin{example}\label{hdns:ex}
  Let $Q = \{\mathsf{q_0,q,q'}\}$, $F = \{\mathsf{q'}\}$.
  The \hdns\ 
  $\mathsf H = \tuple{Q,\mathsf{q_0},\eta,F,\trans}$ where 
  \[\weight{\_} : \begin{cases}
      \mathsf{q_0} \mapsto \{\mathsf{x}\}
      \\
      \mathsf{q} \mapsto \{\mathsf{z}\}
      \\
      \mathsf{q'} \mapsto \emptyset
    \end{cases}
  \qquad\qquad
  \trans : \begin{cases}
    \mathsf{q_0} \mapsto
      \{ \hdtr{\mathsf{x}} {\mathsf{q}} {\sigma:\mathsf{z} \to
      \mathsf{x}} \}
      \\
      \mathsf{q} \mapsto \{\hdtr {\poptr} {\mathsf{q'}} {\bot}, \hdtr {\mathsf{z}} {\mathsf{q}} {\sigma_1:\mathsf{z} \to \mathsf{z}}\}
      \\
      \mathsf{q'} \mapsto \emptyset
    \end{cases}
  \]
  formally defines the \hdns\ in Example~\ref{ex:hdns} (where $\eta$
  is not represented for simplicity).
  \finex
\end{example}

We now define how \hds\ can recognise languages of $\mwords$.
Hereafter, we fix an \hdns
\begin{equation}\label{eq:H}
  \mathcal H \mmdef \tuple{Q,q_0,\eta,F,\trans}
\end{equation}
and, for any stack $\stk$ and any name mapping $\sigma$, we define
$\stkupd \stk \sigma$ by
\[
\stkupd \estk \sigma \mmdef  \sigma :: \estk
\qquad \text{and} \qquad
\stkupd \stk \sigma \mmdef
  \comp \sigma {(\topstk \stk)\upd \fresh \fresh} :: (\popstk \stk)
\]
that basically updates $\stk$ by post-composing its top map $\topstk
\stk$ (if any) with $\sigma$.
Note that this requires $\topstk \stk$ to be updated to allow
composition when $\fresh \in \cod \sigma$.

A \emph{configuration of $\mathcal H$} in~(\ref{eq:H}) is a triple
$\tuple{q,w,\stk}$ where $q \in Q$, $w$ is an $\mwords$, and $\stk$ is
a stack.
Call \emph{initial} a configuration $\tuple{q_0,w,\eta::\estk}$ and
\emph{accepting} $\tuple{q, \emptystr, \stk}$ if $q \in F$.
\begin{definition}\label{def:hdstep}
  Given $q,q' \in Q$ and two configurations $t = \tuple{q, w, \stk}$
  and $t' = \tuple{q',w', \stk'}$, $\mathcal H$ in~(\ref{eq:H})
  \emph{moves from $t$ to $t'$} (written $\step{\mathcal H}{t}{t'}$)
  iff there is $\hdtr \alpha {q'} \sigma \in \trans(q)$ such that
  either of the following cases applies
  \[\begin{cases}
    \alpha \in \weight q
    & \implies
    w = nw' \ \land \
    \topstk \stk(\alpha) = n \ \land \
    \stk' = \stkupd \stk \sigma
    \\
    \alpha = s \in \alp
    & \implies
    w = sw' \ \land \
    \stk' = \stkupd \stk \sigma
    \\
    \alpha = \emptystr
    & \implies
    w' = w \ \land \ 
    \stk' = \stkupd \stk \sigma
    \\
    \alpha = \pushtr
    & \implies
    w' = w \ \land \
    \stk' = \sigma :: \stk
    \\
    \alpha = \poptr
    & \implies
    w' = w \ \land \
    \stk' = \sigma' :: \popstktwo \stk \text{ where }
    \sigma' = \comp{\sigma}{\topstk{\popstk \stk}}
    \\
    \alpha = \oscope
    & \implies
    w = \oscope n. w'\ \land \
    \stk' = \sigma' :: \stk, \text{ where }
    \sigma' = \comp{\sigma}{(\topstk \stk \upd \fresh n)}
    \\
    \alpha = \cscope
    & \implies
    w = \ \cscope w' \ \land \
    \stk' = \sigma' :: \popstktwo \stk, \text{ where }
    \sigma' = \comp{\sigma}{\topstk{\popstk \stk}}
  \end{cases}
  \]

  The set $\rec_{\mathcal H}(t)$ of states reached by $\mathcal H$
  from $t$ on $w$ is defined as
  \[
  \rec_{\mathcal H}(t) \mmdef
  \begin{cases}
    \{q\}
    & \text{if } t = \conf{q, \emptystr, \stk}
    \\
    \bigcup_{\step{\mathcal H}{t}{t'}}{\rec_{\mathcal H}(t')}
    & \text{otherwise}
  \end{cases}
  \]
  A \emph{run of $\mathcal H$ on an m-word $w$} is a sequence of moves
  of $\mathcal H$ from $\tuple{q_0, w, \eta::\estk}$.
\end{definition}
Intuitively, \hdns\ ``consume'' the word in input \emph{moving}
from one configuration to another (likewise classical automata).
However, when the current word starts with a name $n$, the automaton
can progress only if the name ``is known''; namely, it is necessary to
find a transition from the current state $q$ for which the stack maps
a local name of $q$ to $n$.

\hdns\ use a stack ($i$) to keep track of the names of the current
state and, noticeably, ($ii$) to (de)allocate bound names in input
strings.
More precisely, a binder is consumed using a $\oscope$ transition
which updates the meaning of the names.
This is basically done by post-composing the mapping $\sigma$ in the
selected transition with the map on the top of the stack (opportunely
updated to take into account the allocation of $n$).
Instead, a $\cscope$ transition will pop the stack so reassigning
previous meanings to names in the current state by post-composing the
map $\sigma$ of the transition with ``the second one'' in the stack.

\longversion{
  An automaton $\mathcal H$ recognises $w$ if it has a run from its
  initial state to a final state that consumes $w$.
}
\begin{definition}\label{alang:def}
  The \hdns\ $\mathcal H$ in~(\ref{eq:H}) \emph{accepts} (or
  \emph{recognises}) $w$ if $F \cap \rec_{\mathcal H}(\tuple{q_0, w,
    \eta::\estk}) \neq \emptyset$.
  The \emph{language of $\mathcal H$} (written $\lang{\mathcal H}$) is
  the set of words accepted by $\mathcal H$.
\end{definition}
\begin{example}
  If $\mathsf H$ is the \hdns\ in Example~\ref{hdns:ex} and $\eta: x
  \mapsto n$, then $\lang{\mathsf H} = \{n^i \st i>0\}$.
  \finex
\end{example}

Defs~\ref{def:hdstep} and~\ref{alang:def} contain some
subtleties worth spelling out.
First, observe that the language recognised by $\mathcal H$ depends on
$\eta$ which intuitively sets the meaning of the local names of the
initial state $q_0$; instead, the language of $\mathcal H$ does not
depend on the identities of the local names of the states in $\mathcal
H$.
Secondly, an alternative definition would allow the initial stack to
be empty and the correspondence between local names of the states of
$\mathcal H$ and those in the input word is incrementally built during
recognition.
This class of \hdns s would be equivalent to the one in
Defs~\ref{def:hd-automaton} and~\ref{def:hdstep}, but it would have
made our constructions more complex.
Finally, as for classical push-down automata, we could have
equivalently required that an \hdns\ recognises an m-word $w$ only
when it has a run leading to a final state that consumes $w$ and
empties the stack.
We opted for Def~\ref{alang:def} as it is conceptually simpler.
For instance, the following lemma (used to prove
Proposition~\ref{prop:H*}) states that only the top of the stack
is relevant for accepting words.
\begin{lemma}\label{lemma:top}
  Any configuration reachable by an \hdns\ as in~(\ref{eq:H}) from
  $\tuple{q_0,w,\eta :: \estk}$ is also reachable from
  $\tuple{q_0,w,\eta :: \stk}$ for any stack $\stk$.
  \qed
\end{lemma}

In \S~\ref{sec:hsnre} we show how a nominal regular expression $e$ can
be mapped on an \hdns\ $\nretohds e$ that recognises the language of
$e$.
Theorem~\ref{thm:hdnsnre} is the main result
\begin{theorem}\label{thm:hdnsnre}
  For each nominal regular expression $e$, $\lang{\nretohds e} = L(e)$
  interpreted on $\mwords$.
\end{theorem}
\longversion{
  \begin{proof}
   The proof is by induction on the structure of $e$.
   The base cases are trivial while the other cases follow by
   Propositions~\ref{prop:hdsum}, \ref{prop:H1H2}, \ref{prop:H*},
   and~\ref{prop:[n]H}.
   \qed
 \end{proof}
}

\section{\hdns\ and Nominal Regular Expressions}\label{sec:hsnre}
We use nominal regular expressions~(\ref{eq:nre}) to establish a
correspondence between \hdns\ and nominal formal languages.
More precisely, we give (Def~\ref{def:nretohdns}) the map mentioned in
Theorem~\ref{thm:hdnsnre} as the homomorphic image of nominal regular
expression on an algebra of \hdns\ given in the rest of this section.
\begin{definition}\label{def:nretohdns}
   The map $\nretohds{\_}$ from nominal regular expressions to \hdns\
  is defined as:
  \begin{eqnarray*}
  \nretohds 1 & = &
    \tuple{\{q_0,q\},q_0,\bot,\{q\},q_0 \mapsto \{\hdtr \emptystr q \bot \}}
    \text{ where } \weight{q_0} = \weight q = \emptyset
    \\
    \nretohds 0 & = &
    \tuple{\{q_0\},q_0,\bot,\emptyset,q_0 \mapsto \emptyset}
    \text{ where } \weight{q_0} = \emptyset
    \\
    \nretohds n & = &
    \tuple{\{q_0,q\},q_0,x\mapsto n,\{q\},q_0 \mapsto \{ \hdtr x q \bot \}}
    \text{ where } \weight{q_0} = \{x\}, \weight q = \emptyset
    \\
    \nretohds s & = &
    \tuple{\{q_0,q\},q_0,\bot,\{q\},q_0 \mapsto \{ \hdtr s q \bot \}}
    \text{ where } \weight{q_0} = \weight q = \emptyset
    \\
    \nretohds{e_1 + e_2} & = &
    \nretohds{e_1} + \nretohds{e_2}
    \\
    \nretohds{e_1 \circ e_2} & = &
    \nretohds{e_1} \circ \nretohds{e_2}  
    \\
    \nretohds{e^\ast} & = &
    \nretohds{e}^\ast
    \\
    \nretohds{\tuple{n.e}} & = &
    [n] \nretohds{e}
  \end{eqnarray*}
  where the operations on \hdns\ in the last four cases are defined in
  the following.
\end{definition}
The operations on \hdns\ in Def~\ref{def:nretohdns} allow to combine
them so that the language of the resulting \hdns\ has a clear relation
with those the operations act upon as per
Propositions~\ref{prop:hdsum}, \ref{prop:H1H2}, \ref{prop:H*},
and~\ref{prop:[n]H} below.
Theorem~\ref{thm:hdnsnre} can be proved by induction on the structure
of nominal regular expressions using such propositions.
\begin{remark}\label{rmk:nretohdns}
  The map $\nretohds{\_}$ in Def~\ref{def:nretohdns} depends
  on the choice of local names; however, as noted in
  \S~\ref{sec:hdns}, recognisability does not depend on the identity
  of such names.
\end{remark}

The first two clauses in Def~\ref{sec:hsnre} do not involve names and
stack.
Notably, the third clause states that the \hdns\ corresponding to an
expression $n$ has simply a transition from the initial to accepting
state and in the initial configuration the unique name of the former
is mapped to $n$.

The set $\weight{\mathcal H}$ of \emph{(local) names} of an \hdns\
$\mathcal H$ as in~(\ref{eq:H}) is defined as $\weight{\mathcal H}
\mmdef \bigcup_{q \in Q} \weight q$.
In the following, we fix two \hdns
\begin{equation}\label{twohdns:eq}
  \mathcal H_i \mmdef \tuple{Q_i,q_{0,i}, \eta_i, F_i, \trans_i} \text{ for } i \in \{1,2\}
\end{equation}
and, without loss of generality, we assume that $Q_1 \cap Q_2 = \emptyset$
and $\weight{\mathcal H_1} \cap \weight{\mathcal H_2} = \emptyset$.

\begin{definition}\label{def:hdsum}
  Let $q_0 \not\in Q_1 \cup Q_2$ be a new state.
  We define $\mathcal H_1 + \mathcal H_2$ to be the automaton
  $\mathcal H^+ = \tuple{Q^+,q_0^+,\eta^+,F^+,\trans^+}$ where
  \begin{itemize}
  \item $Q^+ = Q_1 \cup Q_2 \cup \{q_0^+\}$ where
    $\weight{q_0^+}_{Q^+} = \weight{q_{0,1}}_{Q_1} \cup
    \weight{q_{0,2}}_{Q_2}$ and $F^+ = F_1 \cup F_2$
  \item $\trans^+(q_0^+) = \{\hdtr \emptystr {q_{0,i}} {id_{\weight{q_{0,i}}}} \st \text{ for } i \in \{1,2\}\}$ and $\trans^+|_{Q_i} = \trans_i$ for $i \in \{1,2\}$, where $id_{\weight{q_{0,i}}}$ is the identity from $\weight{q_{0,i}}_{Q_i}$ to $\weight{q^+_0}_{Q^+}$
  \item $\eta^+ = \eta_1 + \eta_2$, namely $\eta^+(x) = \eta_i(x)$ if
    $x \in \weight{q_{0,i}}_{Q_i}$.
  \end{itemize}
\end{definition}
\begin{proposition}\label{prop:hdsum}
  $\lang{\mathcal{H^+}} = \lang{\mathcal H_1} \cup \lang{\mathcal H_2}$
\end{proposition}
\longversion{
  \begin{proof}
    The statement trivially follows from Def~\ref{def:hdstep} as
    (i)
    $q_0$ has only two outgoing $\emptystr$-transitions which lead
    to the initial states of either of $\mathcal H_1$ or $\mathcal H_2$
    and (ii)
    $\eta$ preserves the name assignments $\eta_1$ and $\eta_2$.
    \qed
  \end{proof}
}

\begin{lemma}\label{lemma:uniquefinal}
  For each \hdns\ $\mathcal H$ there is an \hdns\ $\mathcal H'$ with a
  unique final states and such that $\lang{\mathcal H} =
  \lang{\mathcal H'}$.
\end{lemma}
\longversion{
  \begin{proof}
    Given $\mathcal H$ in~(\ref{eq:H}) and $\hat q \not\in Q$ such
    that $\weight{\hat q} = \emptyset$, we define $\mathcal H' \mmdef
    \tuple{Q \cup \{\hat q\}, q_0, \eta, \{\hat q\}, \trans'}$ where
    $\trans'(\hat q) = \emptyset$, $\trans' = \trans$ when restricted to
    $Q \setminus F$, and $\trans'(q) = \trans(q) \cup \{\hdtr \emptystr
    \hat q \bot\}$ for each $q \in F$.
    The proof that $\lang{\mathcal H} = \lang{\mathcal H'}$ is similar
    to the proof of Proposition~\ref{prop:hdsum}.
    \qed
  \end{proof}
}
Lemma~\ref{lemma:uniquefinal} allows, without loss of generality,
$\mathcal H$ in~(\ref{eq:H}) and each of $\mathcal H_1$ and $\mathcal
H_2$ in~(\ref{twohdns:eq}) to have a single final state, namely $F =
\{q_f\}$, $F_1 = \{q_{f,1}\}$ and $F_2 = \{q_{f,2}\}$, respectively.

The following construction extends the names of an \hdns\ without
altering its language and is used in Def~\ref{def:hdcomp}.
\begin{definition}\label{def:hdaddname}
  Given $\mathcal H$ as in~(\ref{eq:H}) and $x \in \names \setminus
  \weight{\mathcal H}$, $\addname{\mathcal H} x =
  \tuple{Q^\dag,q_0^\dag,\eta^\dag,F^\dag,\trans^\dag}$ is the \hdns\
  such that
  \begin{itemize}
  \item $Q^\dag$ is the named set having the same elements of $Q$ with
    $\weight{\_}_{Q^\dag} : q \mapsto \weight{q}_Q \cup \{x\}$
  \item $F^\dag$ is the named set with the same states of $F$ and
    $\weight{\_}_{F^\dag} : q \mapsto \weight{q}_{Q^\dag} \cup \{x\}$
  \item $\trans^\dag(q) = \{(q',\alpha,\sigma\upd x x) \st
    (q',\alpha,\sigma) \in \trans(q)\}$
  \item $\eta^\dag: \weight{q}_Q \cup \{x\} \to \names$ is the partial
    map undefined on $x$ and behaving as $\eta$ otherwise.
  \end{itemize}
\end{definition}
Hereafter, we assume that $x \in \names \setminus \weight{\mathcal H}$
when writing $\addname{\mathcal H} x$; in fact, by the locality of the
names in the states of an \hdns, if $q$ is a state of $\mathcal H$
such that $x \in \weight q$, we can replace $x$ with any name not in
$\weight q$ by rearranging all the maps in the transitions reaching
$q$.
\begin{lemma}\label{prop:hdaddname}
  $\lang{\addname{\mathcal H} x} = \lang{\mathcal H}$.
\end{lemma}
\longversion{
\begin{proof}
  The proof that $\lang{\mathcal H} \subseteq \lang{\addname{\mathcal
      H} x}$ is trivial as all the transitions of $\mathcal H$ have a
  correspondent in $\addname{\mathcal H} x$ with exactly the same labels
  and name mappings.
  The converse also hold trivially as $x$ cannot play any role in the
  recognition of a word in $\addname{\mathcal H} x$ as $\eta'$ is not
  defined on $x$.
  \qed
\end{proof}
}

\begin{definition}\label{def:hdcomp}
  Let $\{x_1,\ldots,x_j\} = \weight{q_{0,2}}$ and
  $\addname{(\ldots(\addname{\mathcal H_1}{x_1})\ldots}{)x_j} =
  \tuple{Q_1',q_{0,1}',\eta', \{q_{f,1}'\}, \trans'}$.
  The \hdns\ $\mathcal H_1 \circ \mathcal H_2$ is defined as
  $\tuple{Q^\circ,q_{0,1}',\eta^\circ, F_2, \trans^\circ}$ where
  $Q^\circ = Q_1' \cup Q_2$ and
  \[
  \eta^\circ(x) =
  \begin{cases}
   \eta_2(x) & x \in \weight{q_{0,2}}
   \\
   \eta'(x) & \text{otherwise}
  \end{cases}
  \quad
  \trans^\circ(q) = \begin{cases}
    \trans'(q) & q \in Q_1' \setminus \{q_{f,1}'\}
    \\
    \trans_2(q), & q \in Q_2
    \\
    \trans'(q) \cup \{\hdtr \emptystr {q_{0,2}} {id_{\weight{q_{0,2}}}} \},
    & q = q_{f,1}'
  \end{cases}
   \]
\end{definition}
The \hdns\ ${\mathcal H}_1 \circ {\mathcal H}_2$ is built by
connecting the accepting state of ${\mathcal H}_1$ to $q_{0,2}$, the
initial state of ${\mathcal H}_2$, after adding $\weight{q_{0,2}}$ to
${\mathcal H}_1$.
Note that the newly introduced $\emptystr$-transition maintains the
initial meaning of the names in $\weight{q_{0,2}}$ since $\eta^\circ$
acts as $\eta'$ on $\weight{q_{0,2}}$ (and by
Def~\ref{def:hdaddname}).
\begin{remark}\label{rmk:eta}
  A definition more complex than Def~\ref{def:hdcomp} can be given to
  preserve the injectivity of the initial mapping $\eta^\circ$ when
  $\eta_1$ and $\eta_2$ are injective.
  This requires to relax the injectivity condition on $\sigma$ in
  Def~\ref{def:hd-automaton} requiring $\sigma(x) = \sigma(y) \iff
  \sigma(x) = \fresh$ for any $x, y \in \dom \sigma$.
  We opted for the simpler Def~\ref{def:hdcomp} as it just allows more
  non-determism without altering the expressiveness of \hdns.
\end{remark}
\begin{proposition}\label{prop:H1H2}
  $\lang{\mathcal H_1 \circ \mathcal H_2} = \lang{H_1} \circ \lang{H_2}$.
\end{proposition}
\longversion{
  \begin{proof}
    The automaton $\mathcal H_1 \circ \mathcal H_2$ reaches a final
    state iff $w = w_1w_2$ where $w_i \in \lang{\mathcal H_i}$ for
    $i=1,2$.
    In fact, to reach $q_{f,2}$ it is necessary to reach $q_{f,1}$
    first and the unique transition from $q_{f,1}$ to $q_{0,2}$
    maintains on the stack the meaning assigned to the names
    $\weight{q_{0,2}}$ as per the stack.
    \qed
  \end{proof}
}

\begin{definition}\label{def:H*}
  Let $\mathcal H$ be as in~(\ref{eq:H}) with $F = \{q_f\}$.
  The \hdns\ $\mathcal H^\ast = \tuple{Q, q_0, \eta, \{q_f\},
    \trans^\ast}$ is such that
  \begin{eqnarray*}
    \trans^\ast(q) & = & \trans(q),
    \quad\text{ for all } q \in Q \setminus \{q_0, q_f\}
    \\
    \trans^\ast(q_0) & = & \trans(q_0) \cup \{\hdtr \emptystr {q_f} \bot\}
    \\
    \trans^\ast(q_f) & = & \{\tuple{q_0,\pushtr,\eta}\}
  \end{eqnarray*}
\end{definition}
The construction of ${\mathcal H}^\ast$ simply adds an
$\emptystr$-transition from $q_0$ (the initial state of $\mathcal H$)
to $q_f$ (the accepting state of $\mathcal H$) and a
$\pushtr$-transition from $q_f$ to $q_0$ that re-establish the mapping
of the initial configuration preserving in the stack the meaning of
the names.
\begin{proposition}\label{prop:H*}
    $\lang{\mathcal H^\ast} = \lang{\mathcal H}^\ast$
\end{proposition}
\longversion{
  \begin{proof}(Sketch.)
    First, observe that trivially $\emptystr \in \lang{\mathcal H^\ast}
    \cap \lang{\mathcal H}^\ast$ because $\mathcal H^\ast$ has a
    transition $\hdtr \emptystr {\hat q} \bot$ from $q_0$.

    We now prove that $\lang{\mathcal H^\ast} \subseteq \lang{\mathcal
      H}^\ast$.
    If $w \neq \emptystr \in \lang{\mathcal H^\ast}$ then $\mathcal
    H^\ast$ reaches a configuration $\hdtr \emptystr {\hat q} \stk$
    for a suitable $\stk$.
    By construction and Def~\ref{def:hdstep}, $\mathcal H^\ast$
    can visit $\hat q$ only a finite number of times $k$.
    Hence, $w = w_1 \circ \cdots \circ w_k$ where $w_i$ is the word
    processed between the $i$-th visit of $\hat q$ and the previous visit
    of $\hat q$ (or of $q_0$ if $i=1$).
    
    Observing that each visit of $\hat q$ is preceded by a visit of
    $q_f$ (since $\hat q$ can only be reached trough $q_f$), we have
    that $w_1 \in \lang{\mathcal H}$ (and hence in $\lang{\mathcal
      H}^\ast$) because there $q_f$ can be reached from the configuration
    $\tuple{q_0,w_1,\eta::\estk}$.
    For the same reason, we can conclude that $w_{i+1} \in
    \lang{\mathcal H}$ for each $i \in \{1, \ldots, k-1\}$; in fact, the
    $i$-th visit of $\hat q$ yields $\mathcal H^\ast$ in the
    configuration $\tuple{\hat q, w_{i+1} \circ \ldots \circ w_k, \stk}$
    for some stack $\stk$.
    Hence, using the unique transition $\hdtr \pushtr {q_0} \eta$ from
    $\hat q$, the automaton ``resets'' to the configuration
    $\tuple{q_0,w_{i+1} \circ \ldots \circ w_k, \eta::\stk}$, which
    basically amounts to say that $w_i$ can be recognised by $\mathcal
    H$ and the next work $w_{i+1}$ is processed from a configuration
    where $\eta$ is on the top of the stack and the thesis follows
    by Lemma~\ref{lemma:top}.

    We prove that $\lang{\mathcal H}^\ast \subseteq \lang{\mathcal
      H^\ast}$.
    Any word $w \in \lang{\mathcal H}^\ast$ has the form $w = w_1 \circ
    \cdots \circ w_k$ where $w_i \in \lang{\mathcal H}$ for each $i \in
    \{1, \ldots, k\}$, so we proceed  by induction on $k$.
    If $k = 0$ the thesis follows trivially.
    If $k > 0$ then, from the configuration $\tuple{q_0, w_1 \circ w_2
      \circ \ldots \circ w_k, \eta::\estk}$, $\mathcal H^\ast$ reaches a
    configuration $\tuple{q_f, w_2 \circ \ldots \circ w_k, \stk}$ since
    $w_1 \in \lang{\mathcal H}$ by hypothesis.
    Since $\hdtr \emptystr {\hat q} \bot \in \trans'(q_f)$, the
    configuration $\tuple{\hat q, w_2 \circ \ldots \circ w_k, \stkupd
      \stk \bot}$ is reachable from $\mathcal H^\ast$.
    Therefore, $\mathcal H^\ast$ reaches the configuration $\tuple{q_0,
      w_2 \circ \ldots \circ w_k, \eta :: \stkupd \stk \bot}$ which
    yields the thesis by Lemma~\ref{lemma:top}.
    \qed
  \end{proof}
}

\begin{definition}\label{def:[n]H}
  Let $n \in \names$, $\mathcal H$ be as in~(\ref{eq:H}) with $F =
  \{q_f\}$, and let $\hat q, \hat q_f \not\in Q$ be new states with
  $\weight{\hat q} = \weight{q_0} \setminus \eta^{-1}(n)$ and
  $\weight{\hat q_f} = \emptyset$.
  The \hdns\ $[n]\mathcal H = \tuple{Q \cup \{\hat q, \hat q_f\}, \hat
    q, \eta|_{\weight{\hat q}}, \{\hat q_f\}, \trans'}$ is such that
  \[
    \trans'(\hat q)  =  \{\hdtr \oscope {q_0} \sigma\},
    \quad
    \trans'(q)  =  \trans(q),
    \ \forall q \in Q \setminus \{ q_f \},
    \quad
    \trans'(q_f) = \trans(q_f) \cup \{\hdtr \cscope {\hat q_f} \bot\}
  \]
  where $\sigma = \id{\weight{q_0}}\upd x \fresh$, if
  $\eta^{-1}(n) = \{x\}$, otherwise $\sigma = \id{\weight{q_0}}$.
\end{definition}

\begin{proposition}\label{prop:[n]H}
  $\lang{[n]\mathcal H} = [n]\lang{H}$.
\end{proposition}
\longversion{
\begin{proof}
  By construction, $\tuple{\hat q, w, \eta|_{\weight{\hat q}}::\estk}$
  reaches $\hat q_f$ iff there is a word $w'$ such that $w = \oscope n
  . w'$ and $\tuple{q_0, w', \sigma'}$ reaches reaches $\hat q_f$ where
  $\sigma'$ is built as in Def~\ref{def:hdstep}.
  Again by construction, this is possible iff $\tuple{q_0, w',
    \sigma'}$ visits $q_f$ and the last transition which consumes the
  word is a (deallocation) $\cscope$-transition from $q_f$ to $\hat
  q_f$.
  This is equivalent to say that there is $w'' \in \lang{\mathcal H}$
  such that $w' = w''\cscope$ which, by Remark\ref{rmk:nre}, yields
  the thesis.
  \qed
\end{proof}
}

\section{Mapping Nominal Regular Expressions to \hdns}
We build the \hdns\ $\hdof{m(\nrebinder{n.mn})^\ast}$ corresponding
to the expression $m(\nrebinder{n.mn})^\ast$ by applying the
constructions of \S~\ref{sec:hsnre}.
By Definition~\ref{def:nretohdns}, the \hdns\ corresponding to the
expression $m$ is $\hdof m$ with
\begin{equation}\label{eq:Hm}
\nretohds m = \hdof m = \tuple{\{\q_{0,m},\q_{f,m}\}, \q_{0,m}, \eta_m,
  \{\q_{f,m}\}, \trans_m}
\end{equation}
where $\weight{\q_{0,m}} = \{\x\}$, $\weight{\q_{f,m}} = \emptyset$,
$\eta_m \colon \x \mapsto m$, $\trans_m \colon \q_{f,m} \mapsto
\emptyset$, and $\trans_m \colon \q_{0,m} \mapsto \{\hdtr \x {\q_{f,m}}
\bot\}$.
Analogously, the \hdns\ corresponding to the expression $n$ is $\hdof
n$ with
\[
\nretohds n = \hdof n = \tuple{\{\q_{0,n},\q_{f,n}\}, \q_{0,n}, \eta_n,
  \{\q_{0,n}\}, \trans_n}
\]
where $\weight{\q_{0,n}} = \{\y\}$, $\weight{\q_{f,n}} = \emptyset$,
$\eta_n \colon \y \mapsto n$, $\trans_n \colon \q_{f,n} \mapsto
\emptyset$, and $\trans_n \colon \q_{0,n} \mapsto
\{\hdtr \y {\q_{f,n} \bot}\}$.

To compose $\hdof m$ and $\hdof n$, we first have to compute
$\addname{\hdof m} \y$; by Def~\ref{def:hdaddname},
$\addname{\hdof m} \y = \tuple{\Q_\dag,\q_{0,\dag},\eta_\dag,
  \{\q_{f,\dag}\}, \trans_\dag}$ where $\Q_\dag = \{\q_{0,\dag},
\q_{f,\dag}\}$ and
\[
\weight{\_}_\dag \colon
\begin{cases}
\q_{0,\dag} \mapsto \{\x,\y\}
\\
\q_{f,\dag} \mapsto \{\y\}  
\end{cases}
\quad
\eta_\dag \colon \begin{cases}
  \x \mapsto m
  \\
  \y \mapsto \bot
\end{cases}
\quad
\trans_\dag \colon \begin{cases}
 \q_{0,\dag} \mapsto \{\hdtr \x {\q_{f,m}} {\y\mapsto \y}\}
 \\
 \q_{f,\dag} \mapsto \emptyset
\end{cases}
\]

By Def~\ref{def:hdcomp}, $\hdof m \circ
\hdof n = \tuple{\Q_\circ, \q_{0,\dag}, \eta_\circ, \{\q_{f,n}\},
  \trans_\circ}$ where $\Q_\circ = \{\q_{0,\dag}, \q_{f,\dag}, \q_{0,n},
\q_{f,n}\}$,
\[
\eta_\circ \colon \begin{cases}
  \x \mapsto m
  \\
  \y \mapsto n
\end{cases}
\qquad\text{and}\qquad
\trans_\circ \colon \begin{cases}
  \q_{0,\dag} \mapsto \{ \hdtr \x {\q_{f,\dag}} {\y \mapsto \y} \}
  \\
  \q_{f,\dag} \mapsto \{\hdtr \emptystr {\q_{0,n}} {\y\mapsto \y}\}
  \\
  \q_{0,n} \mapsto \{\hdtr \y {\q_{f,n}} {\bot}\}
  \\
  \q_{f,n} \mapsto \emptyset
\end{cases}
\]

We now build $\hdof{\nrebinder{n.mn}} = [n](\hdof m \circ \hdof n)$; let
$\q_s$ and $\q_t$ be two new states with $\weight{\q_s} = \{\x\}$ and
$\weight{\q_t} = \emptyset$, as prescribed by Def~\ref{def:[n]H}, we
have $\hdof{\tuple{n.mn}} = \tuple{\Q_{[n]}, \q_s, \eta_{[n]}, \{\q_t\},
  \trans_{[n]}}$ where $\Q_{[n]} = \Q_\circ \cup \{\q_s, \q_t\}$ and
 the initial setting $\eta_{[n]}$ by restricting
$\eta_\circ$ on $\weight{\q_s}$, i.e.~$\eta_{[n]} \colon \x\mapsto m$;
moreover,
\[
\trans_{[n]} \colon
\begin{cases}
  \q_s \mapsto \{ \hdtr \oscope {\q_{0,\dag}} \sigma \}, & \text{where } \dom \sigma = \{\x,\y\} \text { and } \sigma(\x) = \y  \text { and } \sigma(\y) = \fresh
  \\
  \q_{f,n} \mapsto \{\hdtr \cscope {\q_t} \bot \}
  \\
  \q \mapsto \trans_\dag(\q), & \text{if } \q \in \Q_\dag \setminus \{\q_{f,n}\}
\end{cases}
\]

Further, by Def~\ref{def:H*}, $\hdof{(\tuple{n.mn})^\ast}$ is obtained
by adding two extra transitions $\q_s \mapsto \{\hdtr \emptystr {\q_t}
\bot\}$ and $\q_t \mapsto \{\hdtr \pushtr {\q_s} {\eta_{[n]}}\}$.

Finally, by Def~\ref{def:hdcomp}, we obtain the \hds\ $\hdof
{m\left(\nrebinder{n.mn}\right)^\ast}$ as follows.
First, let $\hdof m ' = \tuple{\{\q_{0,m}',\q_{f,m}'\}, \q_{0,m}',
  \eta_m', \{\q_{f,m}'\}, \trans_m'}$ be obtained as in~(\ref{eq:Hm})
by defining $\weight{\q_{0,m}'} = \{\x'\}$, $\weight{\q_{f,m}'} =
\emptyset$, $\eta_m' \colon \x' \mapsto m$, $\trans_m' \colon \q_{f,m}'
\mapsto \emptyset$, and $\trans_m' \colon \q_{0,m}' \mapsto \{\hdtr {\x'}
{\q_{f,m}'} \bot\}$.
Then, we set $\hdof {m\left(\nrebinder{n.mn}\right)^\ast} = \hdof m '
\circ \hdof{\nrebinder{n.mn}} = \tuple{\Q, \q_{0,m}', \eta, \{\q_t\}, \trans}$
where $\Q = \Q_{[n]} \cup \{\q_{0,m}',\q_{f,m}'\}$ and
\[
\eta \colon \begin{cases}
  \x' \mapsto m
  \\
  \x \mapsto m
\end{cases}
\quad
\trans \colon \begin{cases}
  \q \mapsto \trans_{[n]}(\q), & \text{if } \q \in \Q_{[n]}
  \\
  \q_{0,m}' \mapsto \trans_m'(\q_{f,m}') \cup \{\hdtr \emptystr {\q_{f,m}'} \bot\}
  \\
  \q_{f,m}' \mapsto \trans_m'(\q_{f,m}') \cup \{\hdtr \emptystr {\q_s} {\y \mapsto \y}\}
\end{cases}
\]

  \begin{center}
    \begin{minipage}{\linewidth}\centering
      \includegraphics[scale=.4]{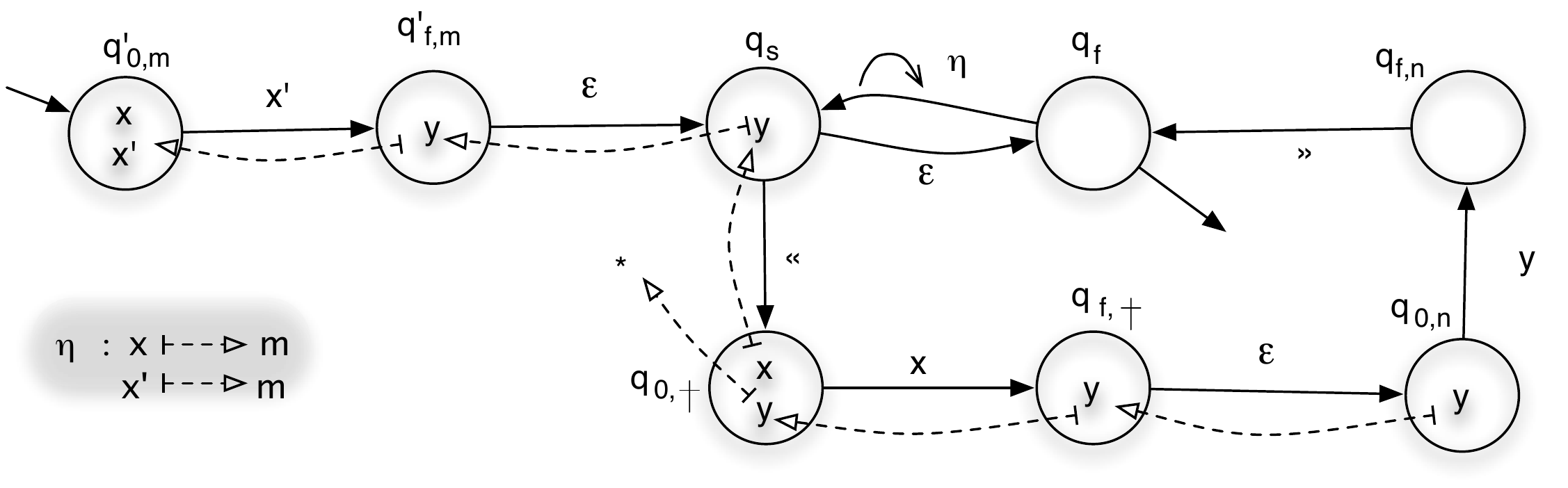}
    \end{minipage}
  \end{center}

We conclude with some final remarks.
Equivalent definitions could have been adopted; for instance, $\eta$
above is not required to be injective (adding some non-determinism in
Def~\ref{def:hdstep}) or some of the new states introduced by the
constructions above could be avoided to obtain more compact \hdns.
We decided to use conceptually simpler constructions instead of more
effective, but more complex ones.

\section{Conclusion}

This paper developed the beginnings of a general theory of words with
binders: nominal languages, nominal monoids, nominal regular
expressions, HD-automata with stacks. We sketch some further work.

Coming back to Table~\ref{tb:axioms} further classes maybe relevant,
for example words satisfying Ax4-5 but not Ax1-3; it will also be of
interest to mix different binders each obeying its own axioms plus
further axioms of their interaction.

HD-automata with stacks are more powerful than necessary if one is
only interested in recognising regular languages; a restricted class
of HD-automata characterising regular languages of m-words can be
described; the same should be done for g-words, l-words, and s-words.

We will also investigate the connections (cf.
Example~\ref{exle:tzev-lang}) of our nominal languages with languages
(on infinite alphabets) without binders~\cite{bojanczyk:stacs11,tze11,gabbayciancia,ct09}.

Further, closure properties and decidability results for these classes
of automata should be studied; for verification purposes deterministic
and minimal automata will be of interest. 

Last but not least, case studies showing the relevance of this line of
research to verification will have to be explored.

\bibliographystyle{abbrv}

\end{document}